\documentclass[twocolumn,epjc3]{svjour3}
\smartqed  
\RequirePackage[T1]{fontenc}
\RequirePackage{mathptmx}      
\RequirePackage{flushend}
\RequirePackage[english]{babel}
\RequirePackage{epsfig}
\RequirePackage{amsmath}
\RequirePackage{amsfonts}
\RequirePackage{amssymb}
\RequirePackage{graphicx}
\RequirePackage{colordvi}
\RequirePackage{color}

\journalname{Eur. Phys. J. C}
\begin{document}

\title{Kinetic theory of Jean Instability in Eddington-inspired Born-Infield gravity}

\author{Ivan De Martino\thanksref{e1,addr1}  \and Antonio Capolupo\thanksref{e2,addr2}}

\institute{Department of Theoretical Physics and History of Science,  University of the Basque Country UPV/EHU,
Faculty of Science and Technology, Barrio Sarriena s/n, 48940 Leioa, Spain\label{addr1}
\and
 Dipartimento di Fisica E.R.Caianiello and INFN gruppo collegato di Salerno,
  Universit\'a di Salerno, Fisciano (SA) - 84084, Italy \label{addr2}}

 \thankstext{e1}{ivan.demartino@ehu.eus}
\thankstext{e2}{capolupo@sa.infn.it}

\date{Received: date / Accepted: date}
\maketitle

\begin{abstract}
We analyze the stability of self-gravitating systems which dynamics is investigated using the collisionless Boltzmann equation, and
the modified Poisson equation of Eddington-inspired Born-Infield gravity. These equations provide a description of the Jeans paradigm 
used to determine the critical scale above which such systems collapse. At equilibrium, the systems are described using the 
time-independent Maxwell- Boltzmann distribution function $f_0(v)$. Considering small perturbations to this equilibrium state, 
we obtain a modified dispersion relation, and we find a new characteristic scale length. Our results indicate that the dynamics 
of the self-gravitating astrophysical systems can be fully addressed in the Eddington-inspired Born-Infield gravity. The latter 
modifies the Jeans instability in high densities environments while its effects become negligible in the star formation regions.
\end{abstract}

\section{Introduction}
\label{uno}

In General Relativity (GR), matter is minimally coupled with the metric and
the  Einstein-Hilbert  Lagrangian,  that is linear in the Ricci scalar, gives rise to the second
order field equations. These are able to explain the dynamics of the particles up to Solar System scale, but
they fail  at scales of galaxies and beyond. The dynamics of self-gravitating systems and the current
period of accelerated expansion of the Universe can not be explained by just the baryonic matter.
Thus, GR needs to incorporate two unknown components to explain the dynamics at both galactic/extragalactic and cosmological scales.
Specifically, almost $\sim68\%$ of the total amount of the matter and energy in the Universe should be in form of
the cosmological constant, or more in general of the Dark Energy, while $\sim26\%$ should be in form of invisible and
exotic particles, named Dark Matter. Nevertheless their fundamental nature is still unknown  \cite{Planck16_13,Feng2010,Khlopov2017}.
The need to incorporate them has been interpreted as a breakdown of GR at astrophysical and cosmological scales
opening to alternative theories of gravity.

Generalizations  of  the gravitational action  have   been  extensively  explored to overcame the need of these two
exotic components. From one hand these are motivated by their capability  to explain the dynamics of self-gravitating systems and
the accelerated expansion of the Universe without resorting to Dark Matter and/or Dark energy
\cite{PhysRept,Annalen,Clifton2012,demartino2014,idm2015,demartino2016,Cai2016,Beltran2017}.
By other hand, they are also motivated by the fact that GR is not the quantum theory of gravity needed to describe the space-time
near the singularities that, as it is well known,  seem can not be avoided \cite{Hawking1973}.
Although a quantum theory of gravity should be able to overcame such problems, it also exists the possibility
to avoid singularities modifying the coupling between matter and gravity. In this context,
Eddington-inspired Born-Infeld (EiBI) gravity has been recently proposed \cite{Banados2010}.
EiBI gravity is inspired to the Born-Infeld action for non-linear electrodynamics, with the Ricci
tensor replacing the field tensor $F_{\mu\nu}$. This structure was motived by some classes of string theories
where the Born-Infeld electrodynamics arises as a low-energy effective theory \cite{Born1934,Fradkin1985}.
One of the most interesting features is that EiBI is equivalent to GR in the vacuum while it introduces
modifications in dense matter environments where GR is experimentally not well probed. EiBI is able to describe,
with only a single extra parameter ($\kappa$), astrophysical objects such as the Sun \cite{Casanellas2012}  and the
internal structure of the compact objects \cite{Pani2011,Pani2012,Avelino2012b,Sham2012,Sham2013,Harko2013,Sotami2014},
and the cosmological expansion of the Universe \cite{Avelino2012b,Avelino2012a,deFelice2012}
(for comprehensive reviews see \cite{Clifton2012,Beltran2017} and reference within).

Briefly, the gravitational action of EiBI gravity
%
takes the following form
\begin{equation}
 S=\frac{2}{\kappa}\int d^4x\biggl(\sqrt{|g_{\mu\nu}+\kappa R_{\mu\nu}|}-\lambda\sqrt{-g}\biggr)+S_{matter}[g,\phi_M],
\end{equation}
where $R_{\mu\nu}$ is the symmetric part of the Ricci tensor, $\phi_M$ represents the matter field, and $\lambda$ is a constant.
The latter is linked to the cosmological constant in such a way that one obtains asymptotically flat solutions setting
$\lambda=1$. Finally, the field equations are built varying the action as in the Palatini approach.
As in other modified theories of gravity, the Palatini's approach
is not equivalent to a pure metric one. However, the latter contains ghosts that can be eliminated only adding extra terms
in the gravitational action \cite{Deser1998,Vollick2004}. The higher order curvature terms account for both non-linear
matter coupling and for avoiding singularities. Such correction terms also appear
in the non-relativistic limit, where EiBI gravity leads to a modified Poisson equation given by
\begin{equation}
 {{\vec \nabla }^2}\Phi (\vec r,t) = 4\pi G\rho (\vec r,t) + \frac{\kappa }{4}{{\vec \nabla }^2}\rho (\vec r,t)\,,
\end{equation}
where $\Phi (\vec r,t)$ is the gravitational potential, and $\rho (\vec r,t)$ is the matter density.
Let us note that by setting $\kappa=0$, the previous equation immediately reduces to the standard Poisson equation
${{\vec \nabla }^2}\Phi (\vec r,t) = 4\pi G\rho (\vec r,t)$.
The tightest  constraint in literature on the EiBI parameter has been obtained comparing the
electromagnetic and gravitational interactions inside atomic nuclei:
$ |\kappa| < 10^{-3} \, {\rm kg^{-1} \, m^5 \, s^{-2}}$ \cite{Avelino2012c}.

In this paper, we  analyze  the kinetic theory of Jean Instability for  self-gravitating systems in
EiBI gravity. This mechanism constitutes, on the theoretical side, a remarkable
instrument to retain/rule out modified theories of gravity at astrophysical level.
In fact,  although a self-gravitating system collapses under the
gravitational force induced by the modified Poisson equation and gives rise to
star formation, we expect that the effects of the EiBI gravity are totally negligible
in star formation environments while they must show some deviations from GR in compact objects collapsing into Black Holes.

The paper is organized as following: in Sec. 2 we compute the dispersion relation for an homogeneous
self-gravitating system in EiBI gravity; in Sec 3 we analyze the dispersion relation in low and high frequency regimes, and
we study the unstable modes that led to the collapse of the structure; in Sec 4 we give our conclusion and remarks.

\section{Dispersion relation of a collisionless self-gravitating system}

The standard approach to describe the collapse of a self-gravitating system, either
a star formation regions such as  interstellar  clouds which physical conditions change
from hot X-ray emitting plasma to cold molecular gas, or compact objects collapsing into Black Hole, is the
Jeans instability \cite{BT}. The latter is usually described  by a distribution function of the particles
$f(\vec r,\vec v,t)$ which is solution of the Boltzman-Vlasov system of equations:
\begin{align}
 & \label{eq:BV1} \biggl[\frac{{\partial}}{{\partial t}} + \left( {\vec v\cdot{{\vec \nabla }_r}} \right) - \left( {\vec \nabla \Phi(\vec r,t) \cdot{{\vec \nabla }_v}} \right)\biggr]f(\vec r,\vec v,t) = I_{coll}\,,\\
 & \label{eq:BV2} {{\vec \nabla }^2}\Phi (\vec r,t) = \biggl[4\pi G + \frac{\kappa }{4}{{\vec \nabla }^2}\biggr]\rho (\vec r, t)\,,
\end{align}
where the mass density distribution reads
\begin{equation}
 \label{eq:BV3} \rho (\vec r,t) = \int {f(\vec r,\vec v,t)} d\vec v\,.
\end{equation}
Here $I_{coll}$ is the collision term and, since we analyze a collisionless system,
it can be neglected ($I_{coll}=0$).

To study the effect of the EiBI gravity,
the eqs. \eqref{eq:BV1}-\eqref{eq:BV3} must be linearized.
Assuming that the unperturbed potential $\Phi_0$ is locally constant in the system,
one can set $\nabla\Phi_0 =0$. This
is generally known as Jeans swindle \cite{BT}.
Thus, in small perturbations regime, the distribution
function and the gravitational potential can be written as
\begin{align}
 & {f(\vec r,\vec v,t) = {f_0}(\vec r,\vec v) + \epsilon{f_1}(\vec r,\vec v,t)}\,,\\
 & {\Phi (\vec r,t) = {\Phi _0}(\vec r) + \epsilon{\Phi _1}(\vec r,t)}\,,
\end{align}
where $\epsilon\ll1$ for small perturbations.
Thus, at the first order, the Boltzman-Vlasov system of equations can be recast in the Fourier space as follows
\begin{align}
 & { - i\omega {f_1} + \vec v\cdot\left( {i\vec k{f_1}} \right) - \left( {i\vec k{\Phi _1}} \right)\cdot\frac{{\partial {f_0}}}{{\partial \vec v}} = 0}\,,\\
&{ - {k^2}{\Phi _1} = 4\pi G\int {{f_1}(\vec r,\vec v,t)} d\vec v - \frac{\kappa }{4}{k^2}\int {{f_1}(\vec r,\vec v,t)} d\vec v}\,,
\end{align}
and the dispersion relation reads
\begin{equation}\label{eq:DispRel1}
{1 + \left( {\frac{{4\pi G}}{{{k^2}}} - \frac{\kappa }{4}} \right)\int {\frac{{\vec k\cdot\frac{{\partial {f_0}}}{{\partial \vec v}}}}{{\vec v\cdot\vec k - \omega }}} d\vec v = 0}.
\end{equation}

Assuming the local thermodynamical equilibrium, the background distribution function of the particles can be described
using the Maxwell-Boltzman distribution. Therefore, $f_0(\vec v)$ is given by
\begin{equation}
 {f_0(\vec v)} = \frac{{{\rho _0}}}{{{{(2\pi {\sigma ^2})}^{\frac{3}{2}}}}}{e^{ - \frac{{{v^2}}}{{2{\sigma ^2}}}}}\,,
\end{equation}
where $\rho_0$ is the matter density at equilibrium, and $\sigma$ is the thermal dispersion velocity of the particles. Finally,
inserting the Maxwell-Boltzman distribution in eq. \eqref{eq:DispRel1} the dispersion relation reads:
\begin{equation}\label{eq:DispRel2}
1 - \left( {\frac{{4\pi G}}{{{k^2}}} - \frac{\kappa }{4}} \right)\frac{{{\rho _0}}}{{\sqrt {2\pi } {\sigma ^3}}} k\int {\frac{{{v}{e^{ - \frac{{v^2}}{{2{\sigma ^2}}}}}}}{{k{v} - \omega }}} d{v} = 0.
\end{equation}
The previous equation shows a singularity at $\omega = kv$. Moreover,  setting $\kappa =0$, it
reduces to the Newtonian dispersion relation.  Thus, one can infer the limit for the collapse
setting $\omega=0$ and computing the maximum wavelength of the perturbations supported by the system,
and above which the system collapses. Specifically, in the Newtonian case, setting $\omega=0$
one obtains the so called Jeans wavenumber
\begin{equation}
 k^2_J = \frac{4\pi G \rho_0}{\sigma^2}\,,
\end{equation}
that can be used to define the Jeans mass as the mass enclosed in a sphere of radius $\lambda_J=2\pi/k_J$, obtaining
\begin{equation}
 M_J = \frac{\pi}{6}\sqrt{\frac{1}{\rho_0}\biggl(\frac{\pi\sigma^2}{G}\biggr)}\,.
\end{equation}
From one hand, perturbations having wavelength $\lambda>\lambda_J$ are unstable, and they exponentially grow. On the other hand,
if the wavelength is less than the Jeans limit then perturbations are strongly damped. In  EiBI gravity,
such limit results to be modified by the additional term in the Poisson equation.
Therefore, from eq. \eqref{eq:DispRel2} we obtain
\begin{equation}\label{eq:EiBIWaveLimit}
 k^{*2}(\omega=0) = \biggl(\frac{\sigma^2}{4\pi G \rho_0} + \frac{\kappa}{16\pi G}\biggr)^{-1}=
 \biggl(1+ \frac{\kappa}{16\pi G}k_J^2\biggr)^{-1} k^2_J\,,
\end{equation}
that corresponds to a mass
\begin{equation}\label{eq:MassLimit1}
M^*={\frac{\pi }{6}\sqrt {\frac{1}{{{\rho _0}}}{{\left( {\frac{{\pi {\sigma ^2}}}{G} + \frac{{4{\pi ^2}{\rho _0}\kappa }}{{16\pi G}}} \right)}^3}} }=
\left(1 + k_J^2 \frac{\kappa}{16\pi G}\right)^{3/2} M_J\,.
\end{equation}
Thus, the mass limit for the collapse of a self-gravitating system depends on
the  EiBI  parameter. It can be higher or lower than the classical Jeans Mass
and, as a consequence, EiBI gravity can favor or disfavor the gravitational collapse
depending on the sign of $\kappa$.

\section{Analysis of the collisionless dispersion relation}

Eq. \eqref{eq:DispRel2} can be easily rewritten as
\begin{equation}
1 - \left( {\frac{{k_J^2}}{{{k^2}}} - \frac{\kappa }{{16\pi G}}k_J^2} \right)\frac{1}{{\sqrt {2\pi } }}\int {\frac{{x{e^{ - \frac{{{x^2}}}{2}}}}}{{\beta  - x}}} dx = 0\,,
\end{equation}
where we have defined the following variables
\begin{equation}
{\beta  = \frac{\omega }{{k\sigma }}}\,;\qquad \qquad {x = \frac{{{v_x}}}{\sigma }}.
\end{equation}
Moreover, in EiBI gravity naturally arises a new wavelength
\begin{equation}\label{eq:lambdaEiBI}
\lambda_{EiBI}=\sqrt {\frac{\pi|\kappa| }{{4G}}}\,,
\end{equation}
that is equal to the one found at cosmological scale and using fluid approach \cite{Avelino2012b}.
The EiBI wavelength allows us to rewrite the dispersion relation in a more compact form
\begin{equation}\label{eq:DispRel3}
1 - \left( {\frac{{k_J^2}}{{{k^2}}} - \frac{k_J^2 }{k_{EiBI}^2}} \right)\frac{1}{{\sqrt {2\pi } }}\int {\frac{{x{e^{ - \frac{{{x^2}}}{2}}}}}{{\beta  - x}}} dx = 0.
\end{equation}

\begin{figure}[!ht]
  \includegraphics[width=\columnwidth]{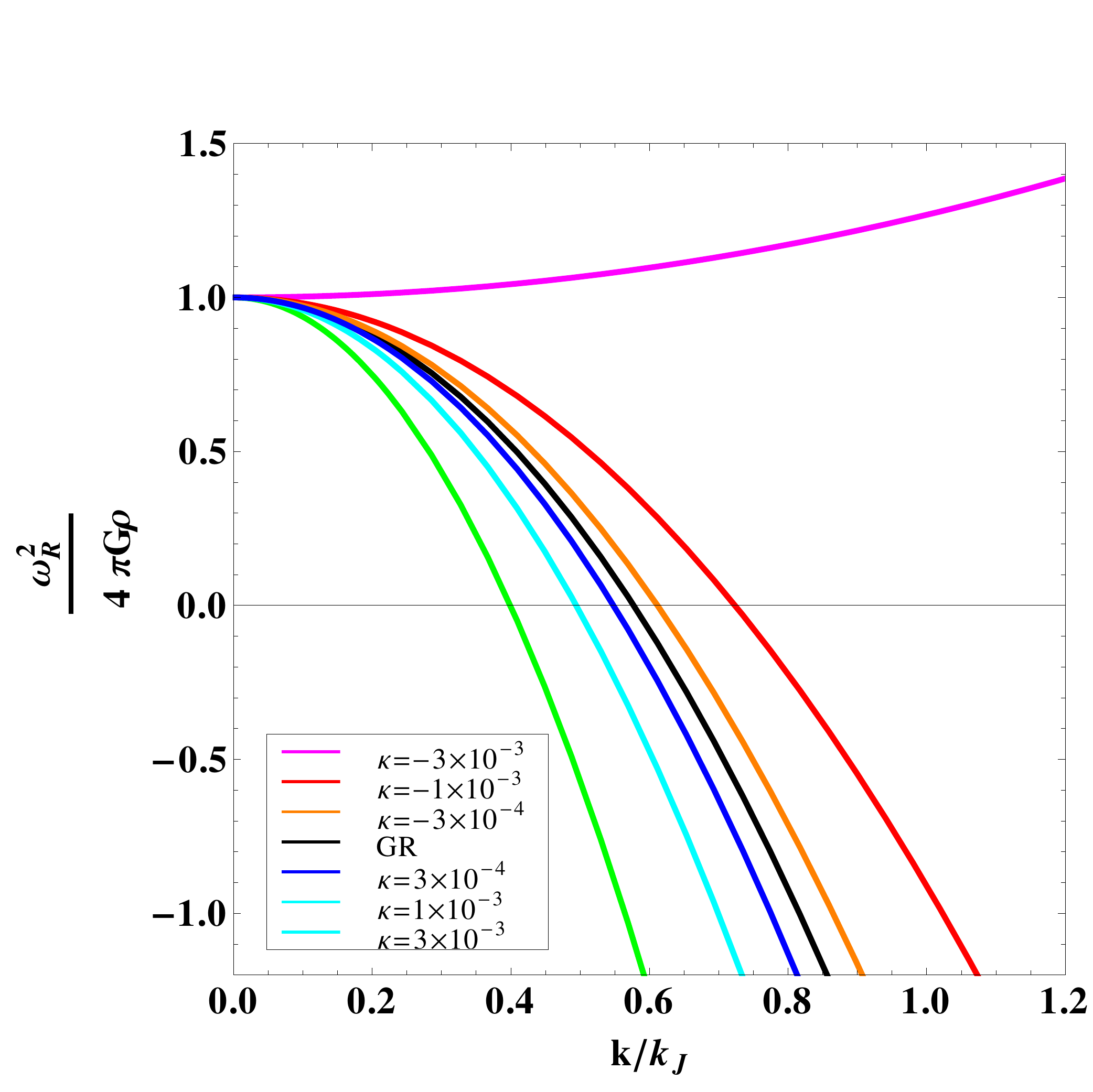}
  \caption{High frequency limit of growth rate of the Jeans instability.
 The oscillation of the plasma are quickly damped (as for the Newtonian case). 
 }\label{fig0}
  \end{figure}

Let us study the limit of high frequency perturbations $\beta\gg1$.
In this case, having no singularities, we can integrate the eq. \eqref{eq:DispRel3}
along the real axis  ($\omega = \omega_R + i\omega_I \approx \omega_R$) obtaining
\begin{equation}
1+\frac{3 k^2 k_J^2 \sigma ^4}{\omega_R ^4}-\frac{3 k^4 k_J^2 \sigma ^4}{k_{EiBI}^2\omega_r ^4}+\frac{k_J^2 \sigma ^2}{\omega_R ^2}-\frac{ k^2 k_J^2 \sigma ^2}{k_{EiBI}^2\omega_R ^2}=0,
\end{equation}
which is a quadratic equation for $\omega_R^2$. Remembering the condition  $\beta\gg1$, we find
\begin{equation}
 \omega_R^2=\biggl[k^2 \biggl(3 +  \frac{k_J^2}{k_{EiBI}^2}\biggr)-k_J^2\biggr] \sigma^2,
\end{equation}
that for $\kappa=0$ reduces to the classical Newtonian relation. 
 Thus, as it is for the Newtonian case, high frequency perturbations with
$k^2 >k_J\left(\frac{k^2_{EiBI}}{3 k^2_{EiBI} + k^2_{J}}\right)$ are quickly damped by the self-gravitating system
whenever the EiBI wavenumber satisfies the condition $k^2_{EiBI} > - k^2_{J}/2$. On the contrary, high frequency perturbations
can be supported by the system as it is shown in Figure \ref{fig0} (magenta line). 

More interesting is to consider the case  of low frequency perturbations $\beta\ll1$
that in Newtonian case led to unstable modes. In such limit, the integral in eq. \eqref{eq:DispRel3} can be recast as
\begin{equation}
 \frac{1}{\sqrt{2\pi}}\int {\frac{{x{e^{ - \frac{{{x^2}}}{2}}}}}{{\beta  - x}}} dx \approx 1 + \imath \sqrt{\frac{\pi}{2}}\beta.
\end{equation}
Therefore,  splitting $\omega$ in its real and imaginary part,  $\omega=\omega_R + \imath \omega_I$, and
setting $\omega_R=0$, we find
\begin{equation}\label{eq:DispRel4}
 \omega_I = k\sigma\sqrt{\frac{2}{\pi}}\biggl[1 - \biggl(\frac{{k_J^2}}{{{k^2}}} - \frac{k_J^2 }{k_{EiBI}^2}\biggr)^{-1} \biggr],
\end{equation}
which gives unstable modes when $\omega_I >0$ that is  for
\begin{equation}
 \biggl(\frac{{k_J^2}}{{{k^2}}} - \frac{k_J^2 }{k_{EiBI}^2}\biggr)^{-1} < 1\,.
\end{equation}
 In other words, the system supports the perturbations having a wavelength
\begin{equation}
 \lambda^2 > \lambda^{*2}\equiv\lambda_{J}^2 - \lambda_{EiBI}^2,
\end{equation}
while it show a singularity for perturbation having $\lambda = \lambda_{EiBI}$.
Perturbations having $\lambda < \lambda^*$ are quickly damped by the system that stay stable.
Next, while in Newtonian gravity perturbations having $\lambda > \lambda_{J}$ are capable to
generate the collapse of the system, in EiBI gravity the limit for the collapse became lower or higher depending on
the parameter $\kappa$. This
can be quickly understood looking at eq. \eqref{eq:MassLimit1}. Since the EiBI modification
depends by the coupling between matter and gravity that is larger as the density increases \cite{Delsate2012},
its effects must be negligible in low density environments like star formation regions where
the condition $\lambda_{EiBI}\ll\lambda_{J}$ must hold. This condition can be straightforwardly translated in to
a density threshold
\begin{equation}
 \rho_0 < \rho^*_0  = \frac{\sigma^2}{\pi^2\kappa},
\end{equation}
which means that EiBI gravity does not affect systems having density below $\rho^*_0$.
From one hand, the interstellar medium has temperatures ranging from 10 to 100 K, while
matter density  ranges from $\rho_0 \sim 10^{-18}$ kg m$^{-3}$ to $\rho_0 \sim 10^{-16}$ kg m$^{-3}$.
Setting $ |\kappa| < 10^{-3} \, {\rm kg^{-1} \, m^5 \, s^{-2}}$ \cite{Avelino2012c} it follows that
the density threshold is $\rho^*_0 \approx 10^7$ kg  m$^{-3}$. As expected,
$\rho_0 < \rho^*_0$ in star formation region. Therefore, EiBI gravity can be neglected.
On the other hand, high density and temperature systems such as
Hyper Massive Neutron Star (HMNS, \cite{Rezzolla2013}) are one
of the most promising laboratory to probe EiBI gravity.
Those systems arise from the merger of a neutron star binary, and
they are expected to collapse to a rotating Black Hole having a massive
accretion torus. The temperature of those systems lies in the range $[5-50]\times 10^{10}$ K
with a number particles density $\sim10^{39}$cm$^{-3}$.
The collapse of such a system is considered as one of the possible source of the short Gamma Ray Bursts.
For such systems,  being $\rho^*_0 \approx 10^{15}$ kg  m$^{-3}$ and
$\rho_0 \approx 10^{18}$ kg  m$^{-3}$,  EiBI gravity affects the kinetic instability
producing deviations from the  GR.

\begin{figure}[!ht]
 \includegraphics[width=\columnwidth]{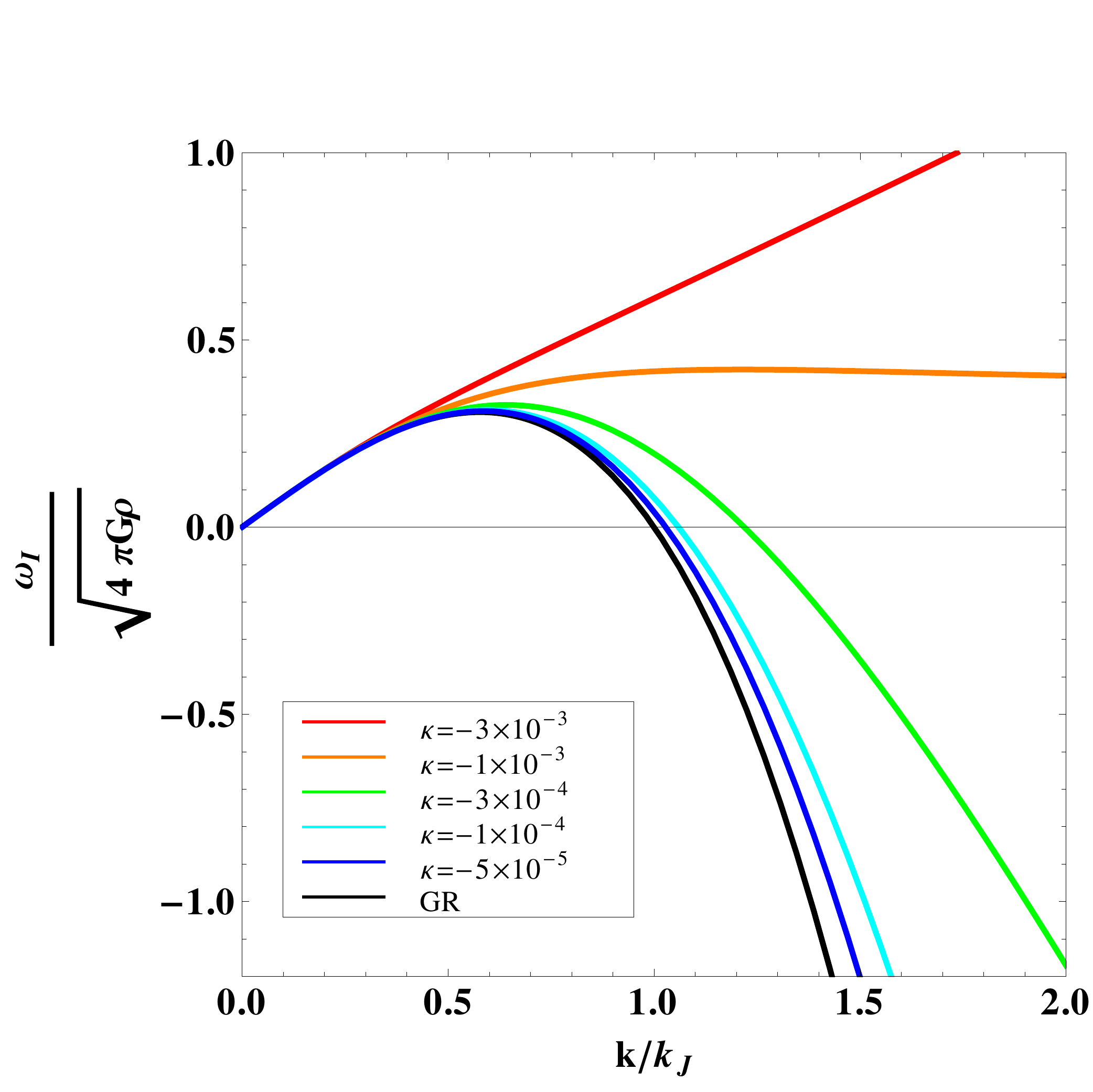}
 \caption{The growth rate of the Jeans instability
versus $k/k_J$ for different values of the EiBI parameter. Here it is illustrated the
dispersion relation for the cases corresponding to $\kappa<0$.}\label{fig1}
  \end{figure}
\begin{figure}[!ht]
 \includegraphics[width=\columnwidth]{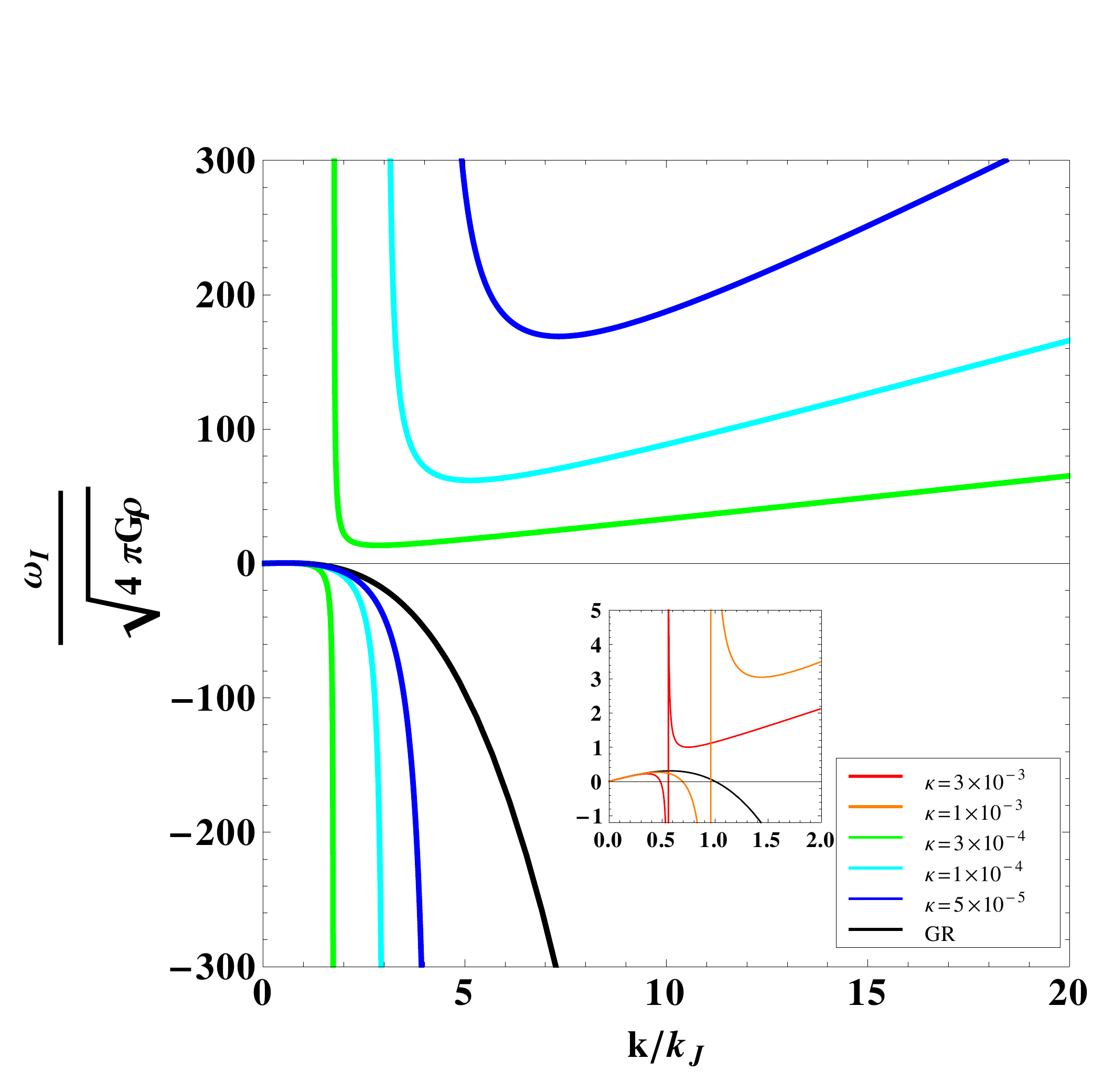}
 \caption{The growth rate of the Jeans instability
versus $k/k_J$ for different values of the EiBI parameter. Here it is illustrated the
dispersion relation for the cases corresponding to $\kappa>0$.}\label{fig2}
 \end{figure}

Figures \ref{fig1} and \ref{fig2} display the growth  rate  of  the  unstable  roots of eq. \eqref{eq:DispRel4}
as function of the normalized  wavenumber $k/k_J$ and for different values of the EiBI parameter $\kappa$. Specifically, Figure \ref{fig1}
illustrates the growth rate for few cases corresponding to negative values of $\kappa$,
while Figure \ref{fig2} is particularized to positive values of $\kappa$.
Figure \ref{fig1} illustrates that the growth rate is larger
for lower values of $\kappa$, while as $\kappa$ is  closer to zero (that corresponds to the Newtonian solution)
the system shows  unstable modes for $k<k^*$, and stable modes for $k>k^*$. Nevertheless,
the Jeans mass results to be higher (see eq. \eqref{eq:MassLimit1}), therefore the
collapse of the HMNS is disfavored with respect to the Newtonian case.

More interesting is the growth rate for positive
values of $\kappa$ shown in Figure \ref{fig2}. In this case the dispersion relation has a singularity at $k=k_{EiBI}$.
The physical behavior around such a point can only be described by more complex physical theories where singularity does not occur.
Therefore, the study here presented shows also the limit of the Jeans instability in EiBI gravity which cannot be blindly applied in the points around $k = k_{EiBI}$.

Notice that such dramatic changes in $\omega_{I}$ corresponds to a phase transition of the system.
In thermodynamic systems,  phase transitions occur when the  free energy, or the partition function
has singularities  for some choice of  variables, such as the temperature.
For example, in the phenomenon of superconductivity, certain materials, when are cooled below a critical temperature $T_{C}$,
pass into the superconducting state characterized by zero electrical resistance and by the complete ejection of magnetic field lines from the interior of the superconductor.
In other systems, other physical parameters play the role of the temperature. For example,  quantum phase transitions
can be obtained by varying the magnetic field or the  pressure at zero temperature.

  In our case, the significant physical parameter is the wavenumber $k$ and the critical value of $k$  which marks the phase
transition is $ k_{EiBI}$.
Indeed, waves having wavenumber smaller  than $k_{EiBI}$ and satisfying the condition $k_<k^*$ give rise to unstable
modes that favor the collapse of the HMNS into a Black Hole. While, waves
having wavenumber higher than the EiBI wavenumber show only stable solutions that do not produce the collapse
of the structure even for $k_{EiBI}<k<k^*$.

We further remark that the extra modes $k \sim k_{EiBI}$ do not exist
in Newtonian gravity where  the singularity in $k = k_{EiBI}$ is absent in such  a model (see Figure 3),
and also that this mode do not depend by the approximation made to compute the dispersion relation in eq.
\eqref{eq:DispRel4} since it is straightforward to highlight that such a singularity is also present
in the general expression of the  dispersion relation given in eq. \eqref{eq:DispRel1}.
Therefore, this sort of phase transition could be an indication that the general paradigm of the Jeans instability, that
works fine in GR and in same other modified gravity models \cite{idm2012},
does not work anymore in EiBI gravity for $k>k_{EiBI}$.


\section{Conclusions and Remarks}

We have investigated the impact of the recently proposed
EiBI gravity theory on the gravitational collapse of a self-gravitating system.
We have solved the collisionless Boltzmann equation  together with the modified
Poisson equation of EiBI gravity to study the kinetic instability
of self-gravitating system, and we have computed the corresponding dispersion relation leading to
a new gravitational scale length. We studied both high and low frequencies limits of the
dispersion relation. In high frequency limit, the self- gravitating system behaves as in Newtonian gravity
not supporting the propagation of the perturbations.
In low frequency limit EiBI gravity may introduce a modification to the Jeans instability.
Although EiBI gravity modified the Jeans Mass needed for the collapse, it does
not affect the star formation because of the low density environments where it happens.
Nevertheless,  in higher density environments such as HMNS the higher order terms of EiBI
gravity produce a departure from  the Newtonian growth rate that could be both an indication that the 
standard Jeans paradigm does not hold anymore, or an effective new modes that could serve as mechanism 
to generate black hole from massive stars.


\section*{Acknowledgements}
IDM acknowledge financial supports from University of the Basque Country UPV/EHU under the program
``Convocatoria de contrataci\'{o}n para la especializaci\'{o}n de personal
investigador doctor en la UPV/EHU 2015", and from the Spanish Ministerio de
Econom\'{\i}a y Competitividad through the research project FIS2010-15492,
and from  the Basque Government through the research project IT-956-16.
AC acknowledge partial financial support from MIUR and INFN.
The authors acknowledges the COST Action CA1511 Cosmology and Astrophysics
Network for Theoretical Advances and Training Actions (CANTATA).


\end{document}